\documentclass[conference]{IEEEtran}
\IEEEoverridecommandlockouts
\usepackage{cite}
\usepackage{amsmath,amssymb,amsfonts}
\usepackage{algorithmic}
\usepackage{graphicx}
\usepackage{subcaption}
\usepackage{textcomp}
\usepackage{xcolor}
\usepackage{url}
\usepackage[export]{adjustbox}
\usepackage{subcaption}
\usepackage{CJKutf8} 
\usepackage{hyperref}
\usepackage{xurl}
\usepackage{float}  % Add this to the preamble for H option
\usepackage{placeins}  % Add this to prevent floats from being misplaced
\usepackage{fancyhdr} % For custom header

\def\BibTeX{{\rm B\kern-.05em{\sc i\kern-.025em b}\kern-.08em
    T\kern-.1667em\lower.7ex\hbox{E}\kern-.125emX}}

\begin{document}
% \maketitle

\title{Enhanced Pediatric Dental Segmentation Using a Custom SegUNet with VGG19 Backbone on Panoramic Radiographs\\
% {\footnotesize \textsuperscript{*}Note: Sub-titles are not captured for https://ieeexplorde.ieee.org  and should not be jjdused}
% \thanks{979-8-3315-1909-4/24/$31.00 ©2024 IEEE}
}

\author{
    Md Ohiduzzaman Ovi\(^1\), Maliha Sanjana\(^1\), Fahad Fahad\(^1\), Mahjabin Runa\(^1\), Zarin Tasnim Rothy\(^1\), \\
    Tanmoy Sarkar Pias\(^2\), A.M.Tayeful Islam\(^1\), and Rumman Ahmed Prodhan\(^1\)\(*\)\thanks{*Corresponding Author: Rumman Ahmed Prodhan (\texttt{rumman.prodhan@nub.ac.bd})} \\
    \(^1\)Department of Computer Science and Engineering, Northern University Bangladesh, Dhaka, Bangladesh \\
    \(^2\)Department of Computer Science, Virginia Tech, Blacksburg, United States \\
    Email: ovi\_41220200210@nub.ac.bd, sanjana\_41210200009@nub.ac.bd, fahad\_41210201589@nub.ac.bd, \\
    runa\_41210200008@nub.ac.bd, rothi\_nubcse@nub.ac.bd, tanmoysarkar@vt.edu, tayef\_nubcse@nub.ac.bd, \\
    rumman.prodhan@nub.ac.bd
}

\maketitle
\thispagestyle{plain}

\begin{abstract}

Pediatric dental segmentation is critical in dental diagnostics, presenting unique challenges due to variations in dental structures and the lower number of pediatric X-ray images. This study proposes a custom SegUNet model with a VGG19 backbone, designed explicitly for pediatric dental segmentation and applied to the Children's Dental Panoramic Radiographs dataset. The SegUNet architecture with a VGG19 backbone has been employed on this dataset for the first time, achieving state-of-the-art performance. The model reached an accuracy of 97.53\%, a dice coefficient of 92.49\%, and an intersection over union (IOU) of 91.46\%, setting a new benchmark for this dataset. These results demonstrate the effectiveness of the VGG19 backbone in enhancing feature extraction and improving segmentation precision. Comprehensive evaluations across metrics, including precision, recall, and specificity, indicate the robustness of this approach. The model's ability to generalize across diverse dental structures makes it a valuable tool for clinical applications in pediatric dental care. It offers a reliable and efficient solution for automated dental diagnostics.
\end{abstract}

\begin{IEEEkeywords}
Pediatric Dental Segmentation, Panoramic Radiographs, Custom SegUNet, VGG19 backbone, Deep Learning in Dentistry, Medical Imaging, AI in Healthcare
\end{IEEEkeywords}

\section{Introduction}
Pediatric dental segmentation is crucial in dental diagnostics and treatment planning, particularly for young patients. However, this task poses significant challenges due to the complexity of pediatric dental structures, the small size of teeth, and the lower quality of pediatric X-ray images compared to adults. These factors make accurate and efficient segmentation difficult, as conventional methods often struggle to handle the variability and subtle details in pediatric dental data.

Deep learning models have shown great promise in medical image segmentation in recent years. However, most studies have focused on adult dental segmentation, leaving a gap in research addressing the unique needs of pediatric patients. Models such as U-Net \cite{p6},\cite{p5} and PSPNet,\cite{p1}, which have been widely used in medical image analysis, have not fully overcome the challenges specific to pediatric dental datasets, particularly about achieving high segmentation accuracy and generalizing across diverse dental structures.

This study proposes a custom SegUNet model utilizing the VGG19 backbone to address these challenges. This model is designed for pediatric dental segmentation and applied to the Children’s Dental Panoramic Radiographs dataset.\cite{p1} To the best of our knowledge, this is the first time the VGG19 backbone has been incorporated into a SegUNet architecture for this dataset. The VGG19 backbone is known for its superior feature extraction capabilities, which enable the model to capture fine-grained details, crucial for accurate segmentation in pediatric dental images.

The proposed model achieves state-of-the-art performance, significantly surpassing existing approaches regarding key segmentation metrics such as accuracy, dice coefficient, and intersection over union (IOU). The model’s ability to generalize across a wide range of pediatric dental structures further highlights its potential for practical application in clinical settings. These results set a new benchmark for pediatric dental segmentation, providing a reliable and efficient solution for automated dental diagnostics.
\\
The key contributions of this study are summarized below:
\begin{itemize}
\item For the first time, a custom SegUNet model with the VGG19 backbone has been applied to the Children’s Dental Panoramic Radiographs dataset.

\item The VGG19 backbone significantly enhanced feature extraction, contributing to the model’s superior segmentation performance.

\item This model achieved state-of-the-art results, surpassing all previous approaches on this dataset.

\item The backbone improved generalization across diverse pediatric dental structures, showing its potential in clinical applications.

\end{itemize}

\section{Literature Review}

In dental X-ray images, multitudinous research has been conducted lately. Zannah et al.\cite{p2} used six U-Net variants on panoramic dental X-ray images, containing a good variety of data. The Vanilla U-Net with two convolutional layers achieved the best performance, with accuracy and IoU. Furthermore, Shuaa S. Alharbi et al.\cite{p3} used Panoramic dental X-ray images obtained from Ivisionlab, each accompanied by a binary mask that isolates the tooth region from the rest of the oral cavity. They used 3 U-Net architectures, where U-Net3+ surpasses others, and were the first to apply the U-Net 3+ model for cavity segmentation. 
   
In addition, Binfeng Zou et al.\cite{p4} propose the DeMambaNet, an innovative segmentation architecture, a cognitively-optimized semantic enhance module, and a hierarchical convergence decoder. They have two types of datasets: The MICCAI teeth segmentation challenge dataset consists of 2D Panoramic dental X-ray images, and the Tufts dental database, which contains 1000 X-ray panoramic radiography images labeled by experts for abnormalities and teeth. DeMambaNet achieved significant performance. Esra Asci et al.\cite{p5} used panoramic radiographs of children aged between 4 and 14, partitioning them into three main categories- primary, mixed, and permanent dentition. Caries lesion segmentation was achieved using a U-Net model implemented in PyTorch; the model demonstrated superior performance in caries lesion segmentation in sensitivity, precision, and F1 score. 
   
Andra Carmen Marginean et al.\cite{p6} proposed CariSeg, an intelligent system trained by applying U-Net architecture, resulting in higher accuracy. CariSeg’s ensemble learning strategy, which incorporates U-shaped neural networks, Feature Pyramid networks, and DeeplabV3, effectively harnesses their complementary strengths. Using UFBA-UESC, Tuft, and pediatric dental radiograph datasets, Suvarna Bhat et al.\cite{p7} custom-built DRA model for binary classification, benefits from a broader dataset, including three separate panoramic dental radiograph datasets, leading to improved training and evaluation across a wider range of caries and abnormalities and periapical abnormalities with accuracies. 
   
Giulia Rubiu et al.\cite{p8} trained an instance segmentation model by applying  (Mask-RCNN) architecture using the Tuft dental database. The model achieved good accuracy and dice coefficient. To address the shortcomings of currently existing methods in tooth segmentation, Jing Hao et al.\cite{p9} proposed TMamba, which includes frequency-based features and shared bi-positional encoding into vision mamba, for improved global feature modeling. TMamba was used to train the TED3 dataset and performed well compared to several state-of-the-art models regarding intersection over union, dice similarity coefficient, and accuracy. In the study of  Emel Gardiyanoğlu et al.\cite{p10}, they aimed to develop an automatic segmentation tool for various objects on orthopantomographs (OPGs). Their dataset had OPG images later converted to PNG files.  
   
Previous studies have not explored using the Children’s dental panoramic radiographs dataset\cite{p1} with SegUnet architecture utilizing the VGG19 backbone.

\section{Dataset}

\begin{figure*}[htbp]  % Use figure* to span two columns
\centering

\vskip\baselineskip

\begin{subfigure}{0.45\textwidth}
    \includegraphics[width=\linewidth]{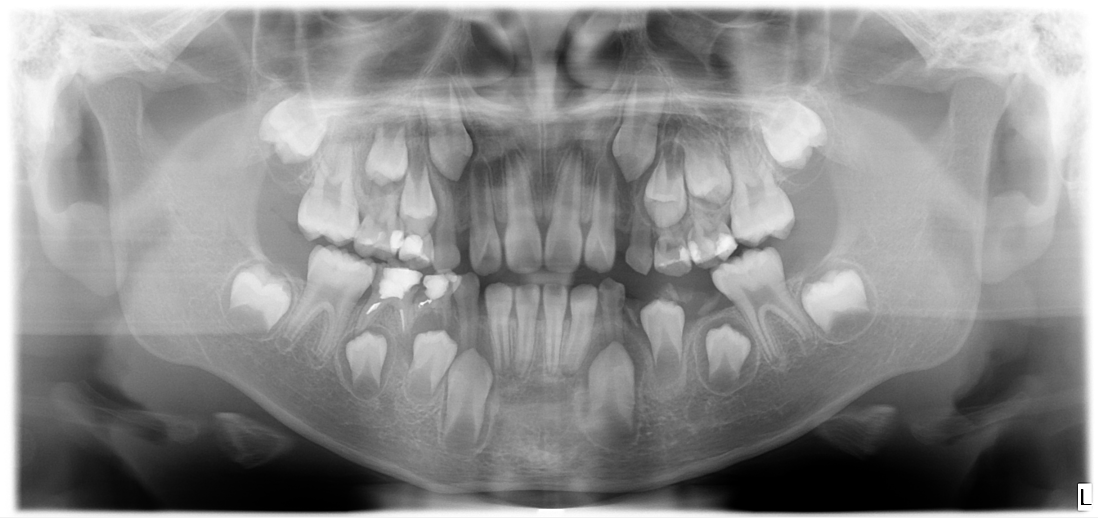}
    \caption{Training Image}
    \label{fig:sub3}
\end{subfigure}
\hfill
\begin{subfigure}{0.45\textwidth}
    \includegraphics[width=\linewidth]{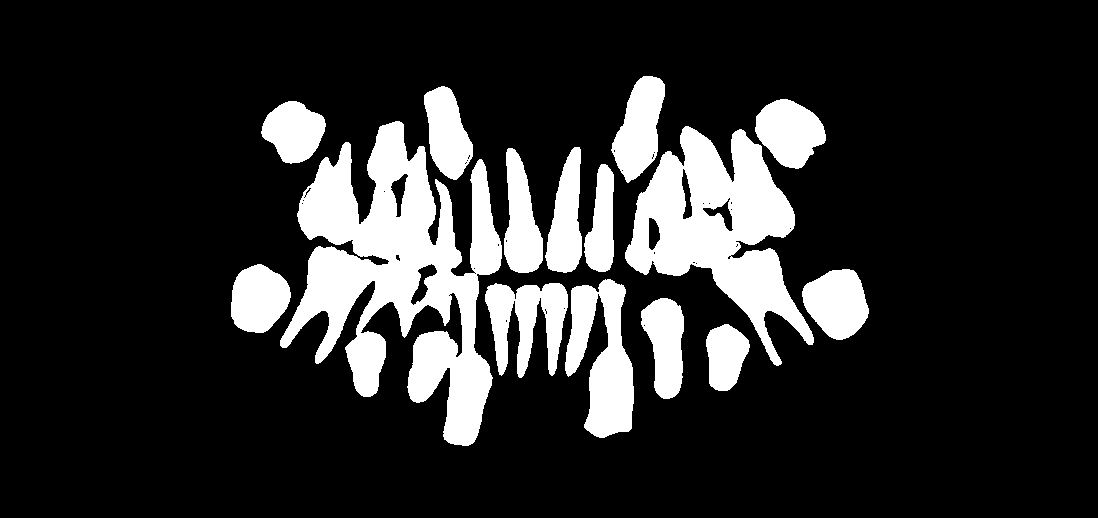}
    \caption{Ground Truth Mask}
    \label{fig:sub4}
\end{subfigure}

\caption{Sample dental panoramic radiograph and mask from Children's Dental Panoramic Radiographs Dataset}
\label{sample_data}
\end{figure*}

\begin{table}[htbp]
\caption{Overview of the Children's Dental Panoramic Radiographs Dataset}
\begin{center}
\begin{tabular}{|p{3cm}|p{5cm}|}
\hline
\textbf{Dataset Attribute} & \textbf{Description} \\
\hline
Dataset Subcategories & 3 (Child Caries Segmentation, Child Disease Detection, Adult Segmentation) \\
\hline
Data Annotations & Segmentation masks (mask files), Test set annotations (JSON files for disease detection) \\
\hline
Children's Tooth Dataset & 193 panoramic X-rays, including 93 of children's teeth (without cases), are provided as supplementary data \\
\hline

New Dataset & 130 panaromic X-Rays\\
\hline

File Formats & png \\
\hline
Types of Data Collected & Panoramic Radiographs (123 films), Intraoral Scans (reference material), CBCT Scans (supplementary material) \\
\hline
\end{tabular}
\label{table1}
\end{center}
\end{table}

The study utilized a publicly available multimodal children dataset\cite{p1}, where the dataset was collected from patient cases at Hangzhou Xiasha Dental Hospital from March through June 2022. The dataset includes 123 films, including panoramic, intraoral, and CBCT scans, with the help of ElSeg and labelMe, an image annotation software. The team collected panoramic X-ray images of 106 patients (ages 2-13). The dataset included three subcategories of dental datasets, including the Children’s Dental Panoramic Radiographs dataset.

The "Child Dental Caries Segmentation Dataset" contains "Train" and "Test" subsets.  with original images, mask files, and JSON annotations. Children's dental caries segmentation dataset includes original and supplement images and the corresponding mask files obtained by segmenting children's dental panoramic X-rays. Also, there was an archive folder with a Chinese name that translates to `New Dataset', and it was later used in this study. Table \ref{table1} provides an overview of the dataset.

The primary datasets (caries segmentation and disease detection) are based on panoramic radiographs because this is the current standard for dental deep-learning studies. Each patient in the dataset has a corresponding panoramic X-ray, and the mask for segmentation is shown in Fig.~\ref{sample_data}.

\begin{figure*}[htbp]  % Use figure* to span both columns
\centering
\includegraphics[width=\textwidth]{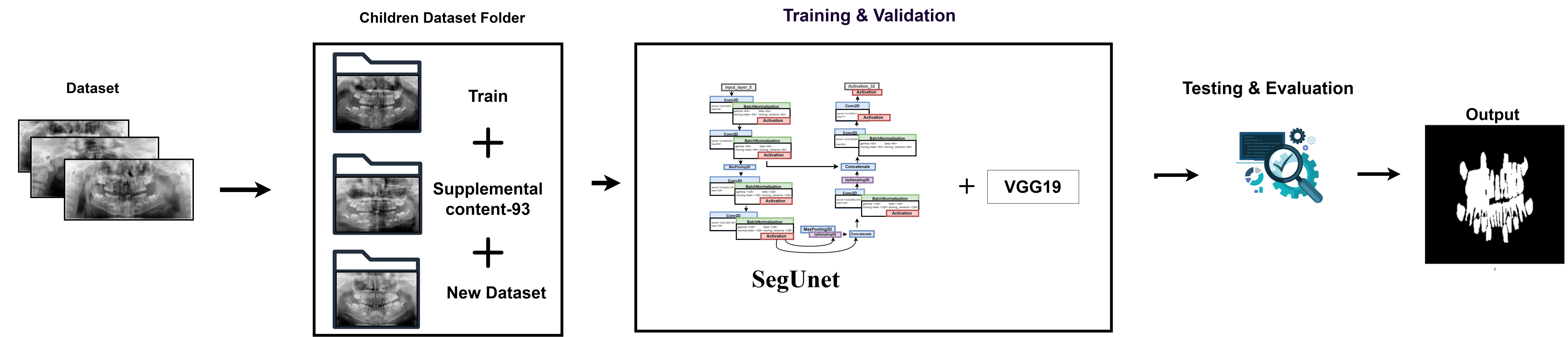}  % Scale to the full text width
\caption{Complete workflow of combining children's radiographs, training the custom SegUNet with VGG19 backbone, and testing on pediatric dental X-rays for segmentation}
\label{workflow}
\end{figure*}

\section{Methodology}\label{AA}
This section outlines the methodology used in this study, focusing on the implementation of the SegUNet architecture with a VGG19 backbone, along with the testing, validation, and evaluation techniques applied to the datasets.

Fig.~\ref{workflow} illustrates the workflow of the study. Data was collected from the Children’s Dental Panoramic Radiographs Dataset, which consists of three folders: Adult Tooth Segmentation, Children’s Dental Caries Segmentation, and Pediatric Dental Disease Detection. Each folder contains Train and Test subsets with original images, mask files, and JSON annotations. Specifically, the Children’s Dental Caries Segmentation Dataset includes the original and supplementary images and the corresponding mask files, which were generated by segmenting children’s dental panoramic X-rays.

In addition to these three primary folders, an archive folder titled \begin{CJK}{UTF8}{gbsn} `新数据集'
\end{CJK} (translated as New Dataset) was also included. This folder, named in Chinese, contains 130 panoramic X-rays and was later incorporated into the study for training and testing purposes. All 323 images and masks were resized to 256 × 256 pixels for uniformity and better processing. Preprocessing steps included loading the images and masks into NumPy arrays and normalizing them by dividing the pixel values by 255.

From the literature, it was noted that most studies using the U-Net architecture \cite{p6}, \cite{p5} and VGG16 backbone \cite{p15} reported good accuracy in dental radiograph segmentation. Therefore, the SegUNet architecture with the VGG19 backbone was chosen for the experiment, adopting the U-Net principles of an encoder-decoder structure, bottleneck layers, and skip connections for improved fine-grained segmentation.

\begin{figure*}[htbp]  % Use figure* to span both columns
\centering
\includegraphics[width=0.7\textwidth, height=1.1\textwidth, keepaspectratio]{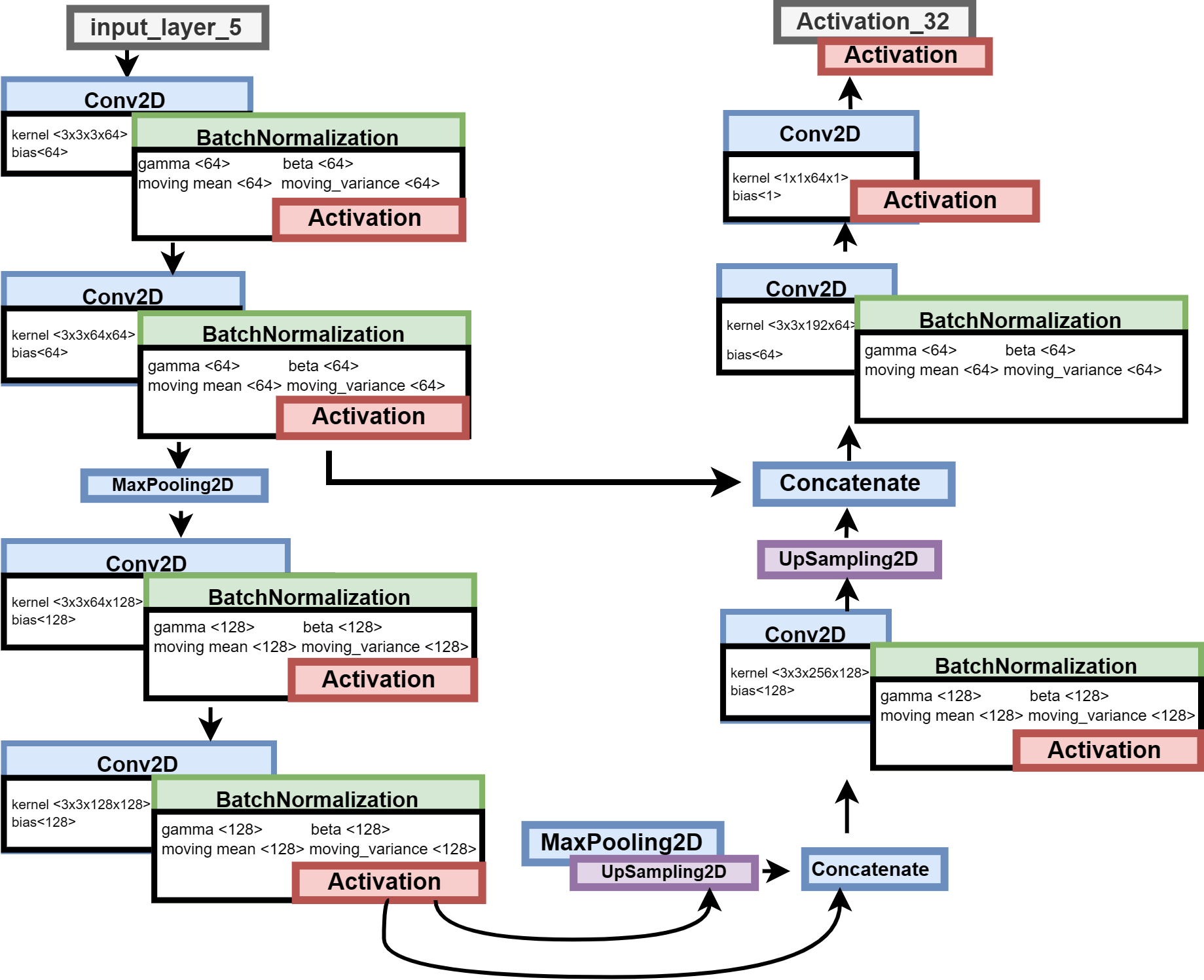}
\caption{Proposed custom SegUNet model architecture}
\label{architecture}
\end{figure*}

The proposed SegUNet architecture shown in Fig. \ref{architecture} is a 13-layer encoder-decoder model designed for semantic segmentation. The encoder starts with two Conv2D layers (64 filters) followed by BatchNormalization and ReLU activation. This is followed by a MaxPooling2D layer to reduce spatial dimensions. This pattern repeats with two more Conv2D layers (128 filters) and another max pooling operation for further downsampling. UpSampling2D layers are used in the decoder to restore the spatial resolution gradually. Concatenate layers bring in skip connections from the encoder to help recover fine details. Each upsampling step is followed by Conv2D layers (128 and 64 filters) with batch normalization and activation. The network ends with a final Conv2D layer and a sigmoid activation for binary segmentation. This structure ensures efficient downsampling and upsampling, preserving important spatial information through the skip connections. The proposed custom SegUNet differs from the vanilla SegUNet by incorporating a pre-trained VGG19 backbone in the encoder, enabling advanced feature extraction and improving segmentation precision. Additionally, BatchNormalization, Dropout, and learnable upsampling with Conv2DTranspose layers enhance generalization and performance on complex datasets.

\section{Experiment and Results}
For the study on the Children's Dental Panoramic Radiographs Dataset, a SegUNet architecture with a VGG19 backbone was implemented to perform segmentation tasks. An extensive literature review revealed that no prior work addressed segmentation on this dataset, except for the original dataset paper. Notably, SegUNet or a backbone was not used in any of the previous experiments.\\
\\
To address this gap, two experiments were conducted using all the data from the Children's dental caries dataset. A train-test split was applied for both the custom SegUNet with a VGG19 backbone and the custom SegUNet without a backbone. These two different setups were trained on the train set from the combined Children's dental caries dataset and tested on the test set. These setups aimed to thoroughly assess the model's performance under different conditions.\\

\subsection{Software and Hardware}

\begin{tabbing}
\hspace{4cm} \= \kill
\textbf{GPU name:} \> NVIDIA L4 Tensor Core GPU \\
\textbf{GPU memory:} \> 24GB \\
\textbf{GPU memory bandwidth:} \> 300GB/s \\
\textbf{Form factor:} \> 1-slot low-profile, PCIe \\
\end{tabbing}
The experiments were conducted using Google Colab Pro, leveraging TensorFlow (v2.17.0)\cite{tf} with the Keras API (v3.4.1
)\cite{keras} as the core framework. Key libraries such as NumPy were utilized for numerical computations, Pandas for efficient data manipulation, Matplotlib\cite{plt} for visualization, and scikit-image[sk]\cite{sk} for image processing. These tools collectively ensured a smooth and effective implementation of the experiments.
\begin{table*}[htbp]
\caption{Performance metrics comparison of segmentation models on the Children’s Dental Panoramic Radiographs dataset}
\label{table2}
\begin{center}
\begin{tabular}{|p{3cm}|p{1.1cm}|p{1.1cm}|p{1.1cm}|p{1.3cm}|p{1.2cm}|p{1.2cm}|p{1.2cm}|p{1.2cm}|}
\hline
\textbf{Model} & \textbf{ACC (\%)}  & \textbf{Dice (\%)}  & \textbf{IoU (\%)}  & \textbf{Recall (\%)}  & \textbf{Precision (\%)}  & \textbf{F1 Score (\%)}    & \textbf{Specificity (\%)} \\
\hline
U-Net \cite{p1} & 97.1\% & 91.2\% & 83.9\% & 92.0\% & N/A & N/A  & 98.0\%  \\
\hline
R2 U-Net \cite{p1} & 96.8\% & 90.3\% & 82.5\% & 88.5\% & N/A & N/A  & 98.4\% \\
\hline
PSPNet \cite{p1} & 96.8\% & 90.8\% & 83.2\% & 88.8\% & N/A & N/A  & \textbf{98.6\%} \\
\hline
Deeplab V3+ \cite{p1} & 96.7\% & 89.6\% & 81.2\% & \textbf{94.9\%} & N/A & N/A  & 97.0\% \\
\hline
Custom SegUnet & 95.9\% & 86.9\% & 86.2\% & 85.1\% & 89.6\% & 87.3\%  & 98.0\%  \\
\hline
Custom SegUNet with VGG19 Backbone & \textbf{97.5\%} & \textbf{92.5\%} & \textbf{91.5\%} & 92.7\% & \textbf{92.3\%} & \textbf{92.5\%}  & 98.5\% \\
\hline
\end{tabular}
\end{center}
\end{table*}

\begin{figure*}[htbp]  % Use figure* to span two columns
\centering
\begin{subfigure}{0.45\textwidth}
    \includegraphics[width=\linewidth]{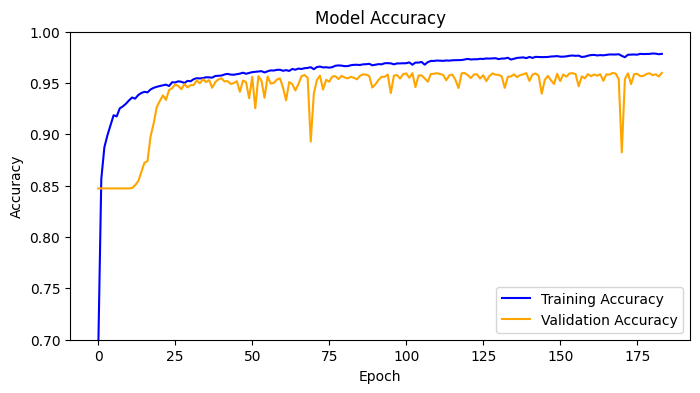}
    \caption{Train and validation accuracy curves for the custom SegUNet\newline}
    \label{fig:sub1acc}
\end{subfigure}
\hfill
\begin{subfigure}{0.45\textwidth}
    \includegraphics[width=\linewidth]{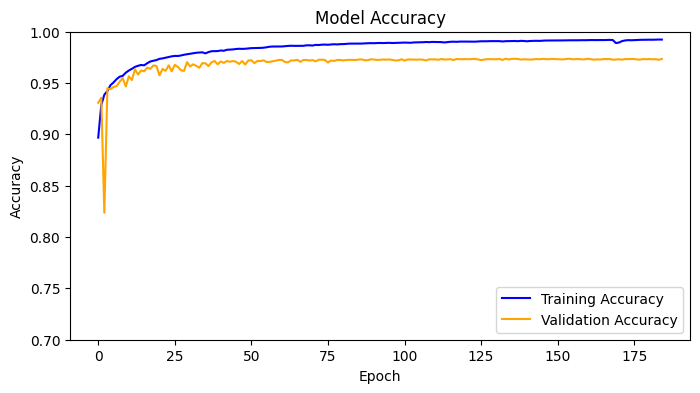}
    \caption{Train and validation accuracy curves for custom SegUnet with VGG19}
    \label{fig:sub2acc}
\end{subfigure}
\vspace{-3mm}  % Reduce the space between rows

\hfill

\caption{Train and validation accuracy curves for the experiments listed on Table \ref{table2}}
\label{accuracy}

\end{figure*}

\subsection{Baseline and Experiment Setup}\label{FAT}
Two different setups were applied. The baseline experiment followed the custom SegUNet architecture shown in Fig. \ref{architecture}, utilizing an encoder-decoder structure for segmentation but without a pre-trained backbone for feature extraction. Based on U-Net principles, this custom architecture served as a baseline for comparing the effects of adding a backbone. The custom SegUNet was enhanced with a pre-trained VGG19 backbone for the second experiment, a model known for its strong feature extraction capabilities. The aim was to evaluate whether this addition would improve segmentation accuracy and overall performance, particularly when handling complex pediatric dental structures.

Both models were trained using the same dataset and experimental setup. The combined dataset was split into 70\% for training, 10\% for validation, and 20\% for testing. All images and their corresponding segmentation masks were resized to 256 x 256 pixels and normalized to ensure consistency. Both models were trained for 200 epochs, with a batch size of 4, using the Adam optimizer with a learning rate of 0.0001. The performance of each model was evaluated on the test set to assess its generalization ability.\\

 \subsection{Performance Metrics}\label{FAT}
 Several key metrics were employed to evaluate the model's performance. Accuracy was used to measure the overall correctness of the predictions. The Dice Coefficient and Intersection over Union (IoU) were particularly important as they indicate the overlap between the predicted and ground truth segmentations, with IoU being a critical metric for assessing segmentation quality. Precision and recall provided insights into the models' ability to correctly identify dental structures, while the F1 Score offered a balanced measure between precision and recall. The Specificity was used to understand further how effectively the models avoided false positives and correctly identified background regions.\\

\subsection{Results and Analysis}\label{FAT}
The experimental results are summarized in Table \ref{table2}, which presents a detailed comparison of the baseline custom SegUNet without backbone and the custom SegUNet with VGG19 backbone, along with other benchmark models such as U-Net, R2 U-Net, PSPNet, and Deeplab V3+. The custom SegUNet with VGG19 backbone achieved the highest accuracy of 97.5\%, significantly outperforming the baseline custom SegUNet without backbone, which achieved an accuracy of 95.9\%, surpassing U-Net's 97.1\%. Regarding the Dice Coefficient, the custom SegUNet with VGG19 backbone scored 92.5\%, demonstrating a notable improvement over the baseline model (86.9\%) and U-Net (91.2\%).

The IoU score for the custom SegUNet with VGG19 backbone was 91.5\%, again outperforming the baseline model (86.2\%) and U-Net (83.9\%), indicating a better overlap between predicted and actual segmentations. The model with the VGG19 backbone also showed superior recall and precision, with recall at 92.7\% and precision at 92.3\%, indicating a strong ability to correctly detect positive cases and avoid false positives. These values were higher than those achieved by the baseline model. Finally, the specificity of the custom SegUNet with VGG19 backbone was 98.5\%, slightly lower than PSPNet’s 98.6\%, but significantly better than the baseline Custom SegUNet, which achieved 98.0\%.

The superior accuracy of the custom SegUNet with the VGG19 backbone compared to the baseline custom SegUNet without a backbone is attributed to the advanced feature extraction capabilities of the pre-trained VGG19. Unlike the vanilla SegUNet, which relies on untrained encoders, the VGG19 backbone captures rich hierarchical features that improve segmentation precision. Additionally, enhancements such as BatchNormalization, Dropout, and robust skip connections further contribute to the model's ability to generalize effectively across diverse pediatric dental structures.

The training and validation accuracy curves (Fig.~\ref{fig:sub1acc} and Fig.~\ref{fig:sub2acc}) provide insight into the performance of the models. In both figures, the training accuracy steadily improves across epochs, showing a strong convergence towards high accuracy. The validation accuracy for the Custom SegUNet model (Fig.~\ref{fig:sub1acc}) stabilizes after an initial increase but shows occasional dips, indicating minor fluctuations in generalization. On the other hand, the Custom SegUNet with VGG19 (Fig.~\ref{fig:sub2acc}) demonstrates a more consistent and stable validation accuracy, aligning closely with the training accuracy and signifying better generalization capabilities with fewer fluctuations. This indicates that incorporating VGG19 enhances model stability and performance on unseen data.

\begin{figure}[htbp]
\centering
\begin{subfigure}{0.48\linewidth}
    \includegraphics[width=\linewidth]{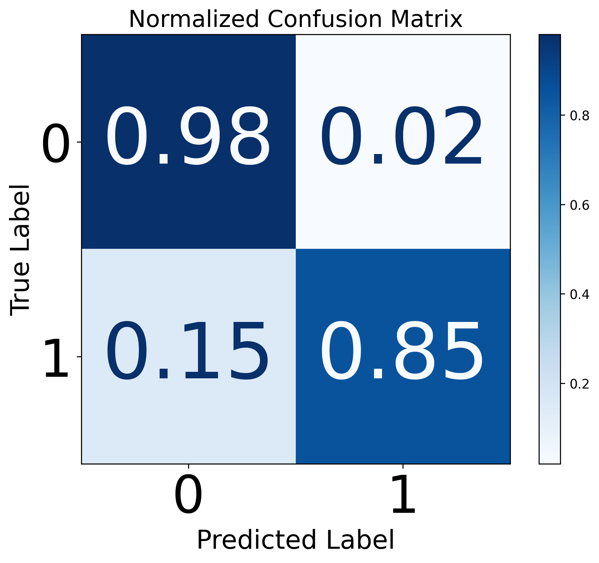}
    \caption{Confusion matrix for custom SegUnet}
    \label{fig:sub1cm}
\end{subfigure}
\hfill
\begin{subfigure}{0.48\linewidth}
    \includegraphics[width=\linewidth]{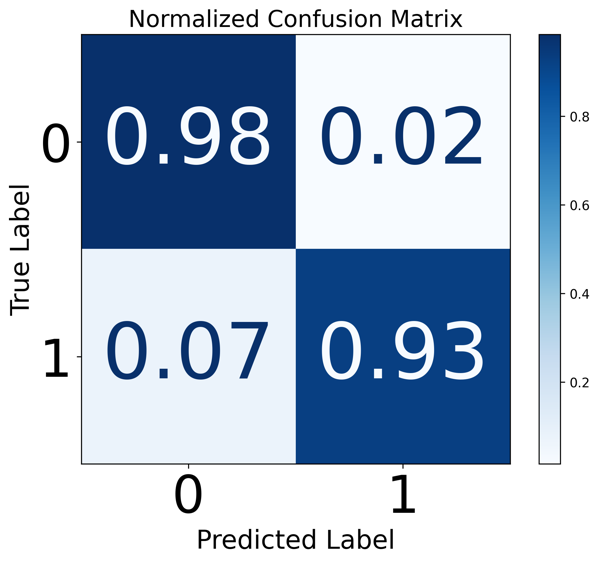}
    \caption{Confusion matrix for custom SegUnet With VGG19}
    \label{fig:sub2cm}
\end{subfigure}
\caption{Normalized confusion matrices for the experiments listed in Table \ref{table2}}
\label{confusion}
\end{figure}

The confusion matrices in Fig.\ref{confusion} illustrate the classification performance for two labels: 0 (background) and 1 (mask). In the Custom SegUNet model (Fig.\ref{fig:sub1cm}), 98\% of the background was correctly classified, whereas the mask was identified with an accuracy of 85\%. In contrast, the Custom SegUNet with a VGG19 backbone (Fig.\ref{fig:sub2cm}) demonstrated superior performance, achieving 93\% accuracy in mask classification and the lowest misclassification rate of only 7\%. This highlights the efficacy of incorporating VGG19 in improving  accuracy.

\subsection{Result Summary}\label{FAT}
The results clearly demonstrate that the custom SegUNet with VGG19 backbone outperformed the custom SegUNet without backbone in all key metrics. Integrating the VGG19 backbone significantly improved the model's feature extraction capabilities, resulting in more accurate and reliable segmentation performance. These findings emphasize the importance of employing pre-trained backbones like VGG19 in complex datasets such as pediatric dental X-rays and set a new benchmark for segmentation tasks in this domain.

\section{Conclusion}
The study introduced a custom SegUNet architecture with a VGG19 backbone for pediatric dental segmentation, applied to the Children’s Dental Panoramic Radiographs dataset. This marks the first known application of the SegUNet and VGG19 combination on this dataset, setting a new benchmark for accuracy and precision in identifying complex dental structures.

Although the results are promising, the relatively small dataset highlights the potential for broader generalizability with expanded and diverse data. Comparing the model with advanced architectures like Vision Transformers \cite{ViT} could also uncover opportunities for further improvements. Additionally, the model's computational complexity and untested performance in clinical settings point to areas for optimization.

Future work could leverage transfer learning from adult dental images, expand the dataset, and integrate clinical feedback during training to enhance practical applications. Exploring advanced or hybrid architectures, as demonstrated in other domains \cite{optimaleeg}-\cite{stroke}, may further improve segmentation accuracy and scalability, paving the way for enhanced clinical adoption.

\section*{Data and Code Availability}
The Children’s Dental Panoramic Radiographs dataset used in this study is publicly available on Figshare \cite{figshare}. The code implementation for the experiments can be accessed on GitHub \cite{code}.

% This will correspond to the reference entry in the reference section.

\end{document}